# Towards Automated Lecture Capture, Navigation and Delivery System for Web-Lecture on Demand

*Rajkumar Kannan and Frederic Andres[1]*

*Department of Computer Science, Bishop Heber College (Autonomous), Tiruchirappalli*
*[1]Digital Content and Media Sciences Research Division*
*National Institute of Informatics, Tokyo, Japan*

**ABSTRACT**

*Institutions all over the world are continuously exploring ways to use ICT in improving teaching and learning effectiveness. The use of course web pages, discussion groups, bulletin boards, and e-mails have shown considerable impact on teaching and learning in significant ways, across all disciplines. ELearning has emerged as an alternative to traditional classroom-based education and training and web lectures can be a powerful addition to traditional lectures. They can even serve as a main content source for learning, provided users can quickly navigate and locate relevant pages in a web lecture. A web lecture consists of video and audio of the presenter and slides complemented with screen capturing. In this paper, an automated approach for recording live lectures and for browsing available web lectures for on-demand applications by end users is presented.*

## INTRODUCTION

The rapid evolution of communication technologies, advanced data networks and semi-automation of multimedia information processing have greatly impacted learning experiences. Multimedia resources include not only entertainment media such as *movie*, *cartoon*, *advertisement*, *computer game*s and so on, but also educational media such as *digital library, museum, digital archives*, and e-learning systems. Lecture recording, like courses and presentations, is one of the important educational media types 1, 2]

The way in which knowledge transfer takes place in a traditional class room-based lecture can, however, be considered highly inefficient. Traditional lectures deliver content in a one-size-fits-all manner to students. It wastes time of more advanced learners by presenting introductory concepts and often is too fast for slow learners who do not possess the required prerequisite knowledge.

Therefore, recording lectures is becoming an increasingly important part of e-learning content generation in several academic institutions. It is proved to be the fast and efficient way of creating content for eLearning 3]. Here a recording system captures most of the information delivered in that lecture and recorded lectures are supplements for a traditional lecture. Apart from being just a kind of e-learning content, it also conveys the feeling that several others are also learning simultaneously and the content is taught by an actual instructor. In other words, viewing a recorded lecture might give a real feeling that can be experienced by all senses, that is users can view the presentation material, see the presenter, and hear his voice as if one is actually attending a physical lecture. Besides, students can watch it wherever and whenever he wants to learn web lectures.

Also, lecture recordings are not only attractive from the learners' point of view. They are also a feasible and cost-efficient way for traditional universities to take first steps into the direction of offering e-learning content 4]. An important factor here is that lecturers do in most cases know how to give a good lecture. For a lecturer, giving a good lecture that is to be recorded does not require significantly much preparation as recording happens automatically.

To fully leverage the possibilities of web lectures, the system must support advanced navigation features such as *skip*, *back*, *keyword search*, *start* and *stop* playing. So that learners must be able to skip unimportant pages of a recording and to repeat problematic sections 5]. This statement is true not only for novel application scenarios but also for conventional lectures when lecture recordings are used as a complement to them. Therefore the system should tackle the problem of time based and structure based navigation in web lectures with a hypermedia navigation concept.

In this paper, we discuss recording of live lectures, their authoring in suitable formats such as *SVG* and *Flex2* and navigation features. The main important contribution of this paper is the introduction of knowledge base in RDF to enable knowledge based queries, apart from the traditional features for navigation in time and structure. The paper is organized as follows: Section 2 surveys related work in lecture capture and delivery systems. Section3 describes the initialization steps to be carried out before recording. Then section4 presents recording and storage of a lecture presentation which includes recording of multimedia slides, video and audio stream of a presenter besides storing them in web servers for later delivery. The different methods of delivering web lectures to end users by applying navigational methods are illustrated in section5. Finally, section6 concludes and presents our next step in the direction of web lectures.

## RELATED WORK

Several approaches consider the problem of automatic lecture capture at the level of single recordings. In these approaches, every step from recording a lecture to creating a fully navigable web lecture is fully automated. Considerable amount of manual work is still necessary for issues such as reservation of lecture halls and distribution of the content produced [6]. Here classrooms are equipped with fixed recording equipments and recording of lectures is carried out with several start/stop sequences. Videos from recording equipments are processed automatically and uploaded to a web server without manual intervention.

The ProjectorBox system seamlessly records RGB information that is sent to a lecture hall's projector and detects when one presentation end and when the next begins using a heuristic method [7]. This system does not record video of the presenter or his handwritten notes and audio signals.

EYA system provides features to record the presenter and audiences who are present in the hall in one-hour interval and stores them on a web server for later delivery [8]. Although these systems completely eliminate the manual intervention in order to prepare and distribute web lectures to end users, they lack features of searching the collection. These systems do not store the details about the content except its title.

The *virtPresenter* lecture recording system [9] tackles navigation in web lectures with a hypermedia navigation concept that is improved with interactive content overviews. Apart from navigation in web lectures the article also addresses didactic scenarios for web lectures and issues related to the workflow of recording lectures. Also, virtPresenter system has been integrated with LMS, thereby the delivery of lecture recordings and access control can thus be facilitated by LMS.

*Stephan et al* [10] applies speech recognition to create a tentative and efficient transcription of the lecture video recordings. These transcription and the words from the power point slides are used to generate semantic metadata serialized in an OWL file. The learning object is defined as the transcription and his videos for one slide. A question-answering system based on these learning objects is presented.

Recently universities have increasingly started to record lectures and offer students these recordings for their learning procedures. Some universities offer even complete series of recordings of hundreds of courses available for public access such as *MIT OCW* [11] and Berkeley courses [12] on YouTube.

Mostly these lecture recordings are indeed multi-modal in nature. They include audio and video streams as well as an information channel for synchronized presentation slides. Additionally, information about the interaction with the presenter PC (e.g. next slide, mouse movements and like) and interaction with other media can add further useful information. Moreover, the linkage with lecture notes, course outline and background information are further information which should be taken into account. Also, teacher-student interaction and student inputs are sources of information to be kept and analyzed 13, 14]

Such media recordings provide a wide range of useful information acquisition within the learning process 15] and also for analysis 16].

CARMUL 17] captures multimodal information where lectures are being scheduled in a classroom in which a teacher and students share the same time and space. The system captures not only hand-writings and slide switching intervals but also audio and video of the people with their spatial location information. These recorded media will be served to users in distance learning process.

A number of smart methods and systems 18] have been proposed to deliver lecture materials including audio, video, and graphical slides for distance learning. These systems consider how to collect interesting or attractive lectures in order to attract attention of end users. Also, some image processing approaches have been proposed 19] for camera control.

The audio and vision analysis approach 20] estimates the location of speakers for a noiseless environment, although this method can estimate the location of multiple speakers precisely.

Since teacher walks around the lecture hall, the trajectory of the teacher is not only a clue to estimate the status of a lecture but also essential information to film him/her properly with pan-tilt-zoom camera. In order to estimate precise trajectory of the teacher, the system integrates location estimation methods by incorporating multiple cameras to estimate 3D location of the teacher in wide area and by analyzing acoustic data using multiple Cross-power Spectrum Phase (M-CSP) analysis 21].

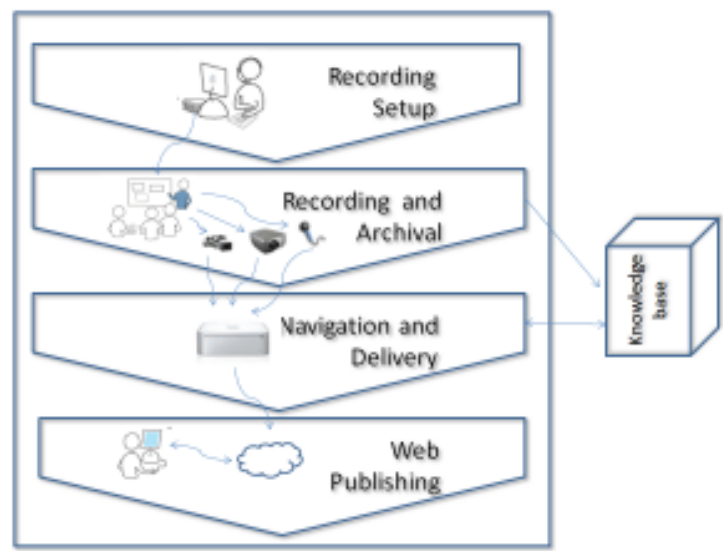

**Fig. 1.** Meta-model of Automated Lecture Capture, Navigation and Delivery System

LectureLounge 22] is a research platform and a system to automatically and non-invasively capture, analyze, annotate, index, archive and publish live presentations. Multimedia Ontology Manager (MOM) is a complete system that allows the creation of multimedia ontology 23]. It supports automatic annotation of lecture videos and question-answering based on automatically-generated AV commentaries and ontology.

## LECTURE RECORDING INITIAL SETUP

Recent advances in recording methods have simplified the recording of audio, video, and image information. Recording lectures in a suitable digital format has several advantages, including the ability to edit lecture content, combine lectures conducted at different times and places, store media digitally, and broadcast conferences electronically over the Internet 24].

The meta-model of our proposed automated lecture capture, browsing and delivery system consists of four components: setup, recording and archiving, navigation and delivery and web publishing as depicted in Figure1. The setup module initiates the process of recording by triggering presenters and students. The recording module carries out audio, video and screen recording that are part of the presentation. Navigation and delivery module provides various features to browse through web lectures and makes them available for web publishing. Finally, web publishing module publishes the encoded web lectures in institution portals, YouTube and iTunesU for public access. More importantly, knowledge base has been designed and can be implemented in RDF to enable knowledge based querying.

The essential components of a slide show are: slide content, slide progression, lecturer's video and voice. Slide content may consist of textual information, images and video clips. During the actual lecture, lecturer defines the slide progression by advancing between slides. To capture slide content and progression, the standard video graphics array (VGA) output from the lecturer's personal computer (PC) is split into two signals. One feed is routed to the digital projector, and the other feed is routed to video converter. An external video camera records the presenter's physical movement in space and his emotions while delivering lecture. The video camera is connected to the application server.

To record audio, an external microphone has been be used in order to improve sound quality and reduce background noise. The output from video converter and video camera are fed to server where the application is running and the application server produces the web lecture in formats such as MPEG4, SVG or Flex2. In order to effectively record live lectures, the application software should allow users to choose different file formats, frame rates and image resolutions. Facilities are needed to adjust the speaker's voice and minimize background noise. Also, recording can be arbitrarily started and stopped based on the lecturer's intension.

## RECORDING AND ARCHIVING LIVE LECTURES

The central administrator can invite a lecturer to perform lecture recording during his lecture presentation. Lecturers can also login to university portal and create a session for live recording. This information is necessary in order to inform students of related disciplines about the addition of this new lecture so as to enable them viewing the web lecture according to their convenience, their learning pace and location for access.

With the video and audio settings properly adjusted, recording of the lecture can be started by the lecturer. The lecturer can start and stop in several sequences the recording according to his convenience several times. Once live lecture is over, the application generates a web lecture in any of the formats and should be archived for later navigation and delivery. Further, compression features are required in order to reduce storage; thereby it occupies reduced bandwidth while streaming the web lecture to PC, iPod or mobile phones over the network.

## NAVIGATION AND DELIVERY OF WEB LECTURES

### Archiving

The low cost of off-line storage options (e.g., external hard drives, optical media) makes archiving a relatively simple and inexpensive task. To enable efficient filtering and rapid browsing of web lectures, we follow unique naming conventions for files such that file names include date, speaker's name and faculty, and an abbreviated title, thereby allowing rapid browsing of available lecture topics.

Our approach to recording multimedia live lectures is both simple and affordable. Another benefit of recorded lectures is the ease of providing access to teaching sessions at multiple sites for distance education. Further, the benefits of web lectures include more efficient use of lecturer's time, whose web lectures are now available to end users at other sites and whose travel time can now be avoided.

### Navigation

Learning will be efficient when both speed in which the learning content is presented and the level at which it is presented is tailored to the learner. For this, web lectures bring the advantage that the material presented in the lecture can be broken up into small pieces that can be replayed as often as required. The possibility to skip known knowledge is the key advantage of eLearning. The basic commonly found navigation features in a slide presentation are timelines and slide overviews. Here timeline provides temporal overview of slides presentation and slide overviews provide structural orientation of slides.

#### *Timeline based navigation*

A number of timeline based navigation approaches exist including random visual scrolling, hierarchical brushing, elastic panning and the Zoomslider. Preview, similar to flip through the book, is possible here by moving the slider knob along the timeline or by brushing over the timeline using the mouse pointer. This allows for a fast assessment of the web lectures.

### Structure based navigation

Structure based navigation is characterized by hypermedia navigations. A hypermedia navigation of web lectures comprises of footprints, bookmarks, full text search, backtracking and structural elements. Footprints highlight visited passages on timeline. Bookmarks are tools which saves important arbitrary passages Full text search indexes all words on the slides that are used during the lecture. Backtracking allows end users to retreat to the previous position so as to undo an undesired target while navigation. Structural elements are arrows which allow jumping to the start of the previous or next animation and slide.

## PERFORMANCE EVALUATION

The high level performance of our proposed system is evaluated with other existing recording and delivery system against a set of features. The evaluative features are characterized based on the principles of recording as well as retrieval. To be more precise, we want the typical web lecture system to provide better ambience of learning and to retrieve the relevant lecture content, apart from user friendly easy access.

**Table 1:** Comparison of Lecture recording and Delivery systems.

|  | Recording of presenter | Recording of audience | Audio recording | Search | Accessibility | Knowledge Discovery |
|---|---|---|---|---|---|---|
| Project Box | No | No | No | No | Low | No |
| EYA | Yes | Yes | Yes | No | Low | No |
| VirPresenter | Yes | Yes | Yes | Yes | Medium | No |
| CARMUL | Yes | Yes | Yes | Yes | Medium | No |
| MIT OCW | Yes | Yes | Yes | Yes | High | No |
| Our System | Yes | Yes | Yes | Yes | High | Yes |

## CONCLUSIONS AND FUTURE WORK

Web lectures are really important sources of eLearning which in these days complement live lectures if not replace them. Several lecture capture and navigation systems in the literature report various ways of recording and few ways of navigation. These systems could be supported with knowledge base, representing various topic elements and their relationships in technologies such as RDF, thereby knowledge assisted search and browsing is realized. This paper introduces knowledge base to identify and archive topics and relationships. Our future research includes the actual implementation of the services required for our proposed system and the design of lecture ontology and studying the performance issues.